# Reduced Dimensionality tailored HN(C)N Pulse Sequences for Efficient Backbone Resonance Assignment of Proteins through Rapid Identification of Sequential HSQC peaks

Dinesh Kumar

Centre of Biomedical Magnetic Resonance, SGPGIMS Campus, Raibareli Road, Lucknow-226014, India

**\*Address for Correspondence:**

Dr. Dinesh Kumar
(Assistant Professor)
Centre of Biomedical Magnetic Resonance (CBMR),
Sanjay Gandhi Post-Graduate Institute of Medical Sciences Campus,
Raibareli Road, Lucknow-226014
Uttar Pradesh-226014, India
Mobile: +91-9044951791,+91-8953261506
Fax: +91-522-2668215
Email: dineshcbmr@gmail.com
Webpage: http://www.cbmr.res.in/dinesh.html

## Abstract:

Two novel reduced dimensionality (RD) experiments –(4,3)D-h<u>NCO</u>caNH and (4,3)D-h<u>N</u>co<u>CA</u>NH– have been presented here to facilitate the backbone resonance assignment of proteins both in terms of its accuracy and speed. The experiments basically represent an improvisation of previously reported HN(C)N experiment [Panchal *et. al.*, J. Biomol. NMR. (2002), 20 (2), 135-147] and exploit the simple reduced dimensionality NMR concept [Szyperski et. al. (2002), Proc. Natl. Acad. Sci. U.S.A. 99(12), 8009-8014] to achieve (i) higher dispersion and resolution along the co-evolved $F_1$ dimension and (ii) rapid identification of sequential HSQC peaks on its $F_2$($^{15}$N)- $F_3$($^1$H) planes. The current implementation is based on the fact that the linear combination of $^{15}$N and $^{13}$C′/$^{13}$C$^\alpha$ chemical shifts offers relatively better dispersion and randomness compared to the individual chemical shifts; thus enables the assignment of crowded HSQC spectra by resolving the ambiguities generally encountered in HN(C)N based assignment protocol because of amide $^{15}$N shift degeneracy. Additionally, each of these experiments enables assignment of backbone $^{13}$C′/$^{13}$C$^\alpha$ resonances as well. Overall, the reduced dimensionality tailored HN(C)N experiments presented here will be of immense value for various structural and functional proteomics studies by NMR; particularly of intrinsically/partially unstructured proteins and medium sized (MW~12-15 kDa) folded proteins. The experiments –like any other experiment that yields protein assignments- would be extremely valuable for protein folding and drug discovery programs as well.

## Introduction:

Backbone amide assignment forms the basis for variety of structural and functional proteomics studies by NMR. Likewise in modern drug discovery research programs, the accurate assignment of $^1$H-$^{15}$N HSQC spectra of target proteins is extremely crucial where backbone amide ($^1$H and $^{15}$N) shift perturbations are used for screening the potential drug candidates [1]. However, the currently available strategies [2-4] based on standard multidimensional NMR experiments [4] –e.g. HNCA/ HN(CO)CA [5,6], HN(CA)CO/HNCO [7], and CBCANH [8]/CBCACONH [9] – are highly time consuming both in terms of data acquisition and data analysis. Most of these strategies involve recording of several 3D spectra and extensive analysis; therefore are neither ideally suited for high-throughput structural and functional proteomics studies by NMR nor for modern drug discovery research programs. Further, the established assignment strategies -where sequential connectivities between the neighboring residues are established using correlated aliphatic $^{13}C^{\alpha/\beta}$ and carbonyl $^{13}C'$ chemical shifts [2,3]- require repeated scanning through the $^{15}$N planes of the 3D spectra to search for the matching correlation peaks along the carbon dimension. Therefore, backbone assignment based on these conventional NMR experiments has remained relatively a slow and arduous process requiring an enormous investment of time and hard-work, particularly in cases when the proteins are intrinsically/partially unstructured or are denatured for various folding/unfolding studies by NMR. In all such cases, unambiguous and rapid identification of sequential HSQC peaks (using these NMR spectra) remains the important rate-limiting step in the resonance assignment process.

In the above context, the pulse sequences proposed earlier for identifying direct sequential correlations between amide ($^1$H and $^{15}$N) chemical shifts of one residue with those of its adjacent neighbors can be very useful [10-12]. However, the strategies based on some of these experiments involve the use two 3D experiments which not only elaborate the analysis, but sometimes may lead to complications due to inter-spectral variations of chemical shifts (a problem common in proteins which are unstable in solution). Interestingly, the HN(C)N experiment proposed earlier [13] and based on novel assignment protocol [14] can be extremely useful in this regard i.e. for rapid sequence specific assignment of backbone $H^N$ and $^{15}$N resonances in ($^{15}$N, $^{13}$C) labeled proteins. The protocol exploits the directly observable amide $H^N$ and $^{15}$N sequential correlations and the distinctive peak patterns in the different planes of the HN(C)N spectrum; the self (intra-residue) and sequential (inter-residue) correlations appear opposite in peak sign (+ and -) and thus can be easily identified without involving any additional complementary experiment. Glycines and prolines, which are responsible for special patterns, provide many check/start points for mapping the stretches of sequentially connected amide cross-peaks on to the primary sequence for assigning them sequence specifically. Overall, the elegant spectral features of HN(C)N enable rapid data analysis and render side chain assignments less crucial for the success of backbone amide assignment. Recently, this novel protocol has also been automated by our group and the algorithm has been given the name AUTOBA [15]. However for a given protein, the success of HN(C)N based assignment protocol depends critically on the dispersion of peaks in its $^1$H–$^{15}$N HSQC spectrum. Overlap of peaks in this spectrum can lead to cancellation of + and - intensities in the HN(C)N spectrum [14]. This happens under two circumstances: first, when the $^{15}$N chemical shifts of neighboring residues are too close and second, when the diagonals having + and - intensities have accidentally close $^{15}$N chemical shifts [14]. In such a situation, the sequential peaks can be absent (because of opposite sign and higher intensity of self peaks in the spectrum) and assignment process may break or may be wrongly considered as proline break point. The situation arises mainly because of the fact that both the indirect dimensions in the HN(C)N spectrum involve $^{15}$N chemical shifts.

In this backdrop, two novel reduced dimensionality experiments –(4,3)D-hNCOcaNH and (4,3)D-hNcoCANH– have been presented here to augment the novel HN(C)N based backbone assignment protocol further, particularly, for protein systems which exhibit high-degree of amide $^{15}$N shift degeneracy. These experiments are a result of a simple modification of the basic HN(C)N pulse sequence [13] and differ in the way the $t_1$ evolution is handled according to reduced dimensionality NMR approach [16,17]. The purpose of current modification is to improve the dispersion and resolution of HN(C)N spectrum mainly to resolve the problems arising because of amide $^{15}$N shift degeneracy. This has been achieved by joint sampling the backbone $^{15}$N$_i$ chemical shift with the backbone $^{13}$C$^\alpha_{i-1}$/$^{13}$C'$_{i-1}$ chemical shift (where *i* is a residue number). This implementation is based on the fact that the linear combination of $^{15}$N$_i$ and $^{13}$C$^\alpha_{i-1}$/$^{13}$C'$_{i-1}$ chemical shifts provides dispersion better compared to the dispersion of the individual chemical shifts as elicited in **Figure S1 (Appendix *I* of Supplementary Material)**. As evident from **Figure S1**, the particular advantage with these new experiments is the dispersion of peaks in the sum and difference frequency regions which can be different; thus facilitates the analysis of the spectrum by simultaneously connecting ($^{15}$N+$^{13}$C) and ($^{15}$N+$^{13}$C) correlations (here $^{13}$C represents backbone $^{13}$C' or $^{13}$C$^\alpha$ chemical shift of sequential *i*-1 residue). However, the only limitation of the method can be long experiment time requirement in order to achieve sufficient resolution along co-evolved NC dimension where spectral width may range from 2 to 3 times the original $^{15}$N spectral width. Thus for identical resolution, the total experiment time in each case will be almost two-three times to that of the basic HN(C)N experiment. However, for proteins with high shift degeneracy, the benefit incurred in terms of increased spectral resolution and dispersion becomes more crucial compared to the increased total experiment time.

Nevertheless, the longer experiment time problem can be circumvented by using non-uniform (sparse) sampling along indirectly detected dimensions [18]; thus making NMR data acquisition fast. Since the experiments proposed here does not involve aliphatic protons, the overall experiment time can be reduced further by modifying these experiments according to either BEST NMR [19] or L-optimization [20] approaches which allow rapid data collection of NMR spectra via minimizing the inter-scan recycle delay without loss of sensitivity. Rapid data collection generally becomes a crucial issue whenever the protein under investigation is unstable in solution; the protein either precipitates or degrades inside the NMR sample tube in matter of days. The other disadvantage associated with the experiments presented here is their slightly reduced sensitivity compared to the basic HN(C)N experiment. The sensitivity will be lower (a) by a factor of 2 with respect to the basic HN(C)N experiment and (b) by a factor of $\sqrt{2}$ compared to an alternative 4D experiment with separate frequency labeling (e.g. like 4D-hNCOCANH). However with the currently available higher magnetic fields and efficient cryo-probes offering higher signal to noise ratio, the sensitivity of the experiment will not be a serious issue and both these experiments can be successfully applied on higher (< 15 kDa) molecular weight proteins; particularly the intrinsically unstructured or chemically denatured proteins where the inherent backbone flexibility renders increased sensitivity of the experiment [21,22]. Further, both these experiments provide easy identification of self and sequential correlation peaks because of their opposite signs; these are readily amenable to be exploited in automated assignment algorithms like any other set of NMR experiments that yield protein assignments in an automated manner [21]. Performance of these experiments has been tested and demonstrated here on two folded proteins -bovine apo calbindin-d9k ($Ca^{2+}$ free apo form) and human ubiquitin- and on a highly disordered protein UNC60B denatured in 8M urea.

**Materials and Methods:**

The proposed reduced dimensionality experiments -(4,3)D-hNCOcaNH and (4,3)D-hNcoCANH- have successfully been tested and demonstrated on three proteins including both folded (e.g. 75 amino acid long bovine apo Calbindin-d9k and 76 amino acid long human ubiquitin) and unfolded ones (e.g. unfolded UNC60B, a 152 amino acid protein denatured in 8M urea). The $^{13}$C/$^{15}$N labeled bovine apo Calbindin sample (1 mM in concentration, 50 mM Ammonium Acetate, pH 6.0, 10% D$_2$O, in high quality NMR tube sealed under inert atmosphere) has been purchased from Giotto Biotech, Itlay: *http://www.giottobiotech.com/*); whereas $^{13}$C/$^{15}$N labeled human ubiquitin (1.2 mM dissolved in phosphate buffer pH 6.5 in 90% H$_2$O and 10% D$_2$O has been purchased from Cambridge Isotope Laboratories, Inc., USA (*http://www.isotope.com*). The $^{13}$C/$^{15}$N labeled sample (final concentration ∼ 0.8 mM and pH 6.0) of 8M urea-denatured state of UNC60B (a 152 amino acid ADF/cofilin family protein of *Caenorhabditis elegans*) has been prepared as described in **Appendix *II*** (Supplementary Material). All the experiments have been performed at 298 K on a Bruker Avance III 800 MHz NMR spectrometer equipped with a cryoprobe. The various acquisition and processing parameters used in these experiments have been listed in **Table S1.** The delays $2T_N$, $2\tau$, and $2\tau_{CN}$ were set to 28, 9, and 25 ms, respectively. Gaussian cascade Q3 pulses [23] were used for band selective excitation and inversion along the $^{13}$C channel. The $^{13}$C carrier frequencies for pulses in $^{13}$C$^\alpha$ and $^{13}$C' channel were set at 54.0 ppm and 174 ppm, respectively. WATERGATE scheme [24] is used to suppress the water signal. Water saturation is minimized by keeping water magnetization along the z-axis during acquisition with the use of water selective 90º pulses [25]. The NMR data was processed using Topspin (BRUKER, http://www.bruker.com/) and analyzed using CARA [26]. A point to be mentioned here is that though the experiments have been carried out at 800 MHz, the high magnetic fields can be detrimental for the sensitivity of these experiments. This is because of the fact that the transverse relaxation rate of the carbonyl carbons is dominated by the chemical shift anisotropy mechanism [27] and both these experiments involve substantially long delays with transverse carbonyl carbon magnetization required for the $^{13}$C'→$^{13}$C$^\alpha$ transfers **(Fig. 1A and 1B)**.

## Results and Discussion:

### *Description of Pulse sequences*

The schematic presentation of magnetization transfer pathway along with the respective frequency labeling schemes and the pulse sequences of the proposed reduced dimensionality experiments –(4,3)D-hNCOcaNH and (4,3)D-hNcoCANH– have been shown in **Fig. 1**. **Figure 1A and 1B** traces the magnetization transfer pathways and **Fig. 1C and 1D** represent, respectively, the pulse sequences –(4,3)D-hNCOcaNH and (4,3)D-hNcoCANH. The pulse sequences have been derived by the simple modification of the basic 3D HN(C)N experiment described earlier [13] and differ in the way the $t_1$ evolution is handled according to reduced dimensionality (RD) NMR [16,17]. In case of (4,3)D-hNCOcaNH experiment, the $t_1$ evolution period involves co-evolution of backbone $^{15}N_i$ and $^{13}C'_{i-1}$ chemical shifts **(Fig. 1A and 1C)**; while in case of (4,3)D-hNcoCANH experiment, the $t_1$ evolution involves co-evolution of backbone $^{15}N_i$ and $^{13}C^{\alpha}_{i-1}$ chemical shifts **(Fig. 1B and 1D)**. Each experiment leads to the spectrum equivalent to basic HN(C)N spectrum, but with higher chemical shift dispersion and randomness along the $F_1$ axis, referred here as NC dimension. Therefore, the transfer efficiency functions which dictate the intensities of self and sequential correlation peaks in their respective spectra would be the same as those in the HN(C)N spectrum described previously [13,28,29]. Therefore, only the distinguishing features of these spectra have been presented here. As shown schematically in **Figure 2**, the peaks appear at the following coordinates in their respective spectra:

$$F_1 = N_i^+ / N_i^-, \ (F_2, F_3) = (H_i, N_i), (H_{i-1}, N_{i-1})$$
$$F_2 = N_i, \ (F_1, F_3) = (H_i, N_i^+), (H_i, N_{i+1}^+), (H_i, N_i^-), (H_i, N_{i+1}^-)$$

The letters "H" and "N" here refer to amide $^1H$ and $^{15}N$ chemical shifts, whereas the letters "$N_i^+$" and "$N_i^-$" refer to the linear combination of $^{15}N_i$ and $^{13}C_{i-1}$ chemical shifts (where $i$ is residue number and $^{13}C$ represents the backbone $^{13}C^{\alpha}$ or $^{13}C'$ chemical shift). In **Figure 2**, the letters $N^+$ and $N^-$, have been represented as $^{15}N^+$ and $^{15}N^-$, respectively, and depending upon the offset of RF pulses used along $^{13}C$ channel (i.e. $^{13}C_{offset}$), these can be evaluated as:

$$^{15}N_i^+ = {^{15}N_i^{obs}} + \kappa*({^{13}C_{i-1}^{obs}} - {^{13}C_{offset}}) \quad (1)$$
$$^{15}N_i^- = {^{15}N_i^{obs}} - \kappa*({^{13}C_{i-1}^{obs}} - {^{13}C_{offset}}) \quad (2)$$

where "k" is the scaling factor and is equal to $\gamma(^{13}C)/\gamma(^{15}N)=2.48$.

Thus, the $F_2(^{15}N)$-$F_3(^1H)$ plane corresponding to $F_1 = N_i^+ / N_i^-$, shows two intra-residue amide correlation peaks: a self ($H_i$, $N_i$) and a sequential ($H_{i-1}$, $N_{i-1}$) **(Fig. 2, right panel)** and the $F_1(NC)$-$F_3(^1H)$ plane corresponding to $F_2 = N_i$, shows four correlation peaks: $(H_i, N_i^+, H_i, N_i^-, H_i, N_{i+1}^+ \text{ and } H_i, N_{i+1}^-)$ **(Fig. 2, left panel)**. As shown in Figure 1, the self (notion used here for $H_i N_i$, $N_i^+$ and $N_i^-$) and sequential (notion used here for $H_{i+1} N_{i+1}$, $N_{i+1}^+$ and $N_{i+1}^-$) correlation peaks have opposite peak signs in different planes of these spectra except in special situations; in situations where a Glycine/Proline residue is involved, the actual sign patterns of the two kinds of peaks in each set would vary as in HN(C)N [10]. For glycines, different patterns arise depending on whether the $i$th or $i$-1th residue is a glycine or otherwise. For example, in a stretch –XGYZ-, in the $F_1$-$F_3$ plane at the $^{15}$N chemical shift of G, all peaks appear with same sign (positive) and likewise, at the $^{15}$N chemical shift of Y, all the four peaks will also have the same but opposite sign (i.e. Negative). For the same reason, in a double glycine stretch like GG'Z the inter- and intra-residue correlation peaks have distinct patterns. Similarly, in the case of proline at position $i$+1, absence of amide proton results in the absence of sequential ($H_i, N_{i+1}^+$ and $H_i, N_{i+1}^-$) correlation peaks. These special peak patterns at glycines, proline and the residues present next to them provide important start or check points during the course of sequential assignment process. Thus, for a given protein with a known amino acid sequence, it is possible to identify several special triplet sequences simply by inspecting the various $F_1$–$F_3$ and $F_2$-$F_3$ planes of these spectra. Considering different triplets of residues, covering the general and all the special situations, all the expected peak patterns in the $F_1$-$F_3$ and $F_2$-$F_3$ planes of these spectra are schematically shown in **Figure S2** (with details explained in **Appendix-III**, Supplementary material).

The above described features of the reduced dimensionality HN(C)N experiments have been tested and demonstrated here on all the three proteins: bovine apo calbindin (75 aa), human ubiquitin (76 aa) and unfolded UNC60B (152 aa). **Figure 3** shows the experimental demonstration of these features for the $F_1$(NC)-$F_3$($^1$H) and $F_2$($^{15}$N)-$F_3$($^1$H) planes of the (4,3)D-hNcoCANH spectrum of unfolded UNC60B. **Figure S3A** and **S3B** shows the experimental demonstration of these features for the $F_1$(NC)-$F_3$($^1$H) planes of (4,3)D-hNCOcaNH spectrum of human ubiquitin and (4,3)D-hNcoCANH spectrum of bovine apo-calbindin, respectively. **Figure S4** shows the experimental demonstration of these features for the $F_2$($^{15}$N)-$F_3$($^1$H) planes of the (4,3)D-hNcoCANH spectrum of bovine apo-calbindin.

### Facile Identification of Sequential Amide-Cross Peaks:

Like basic HN(C)N experiment, the beneficial feature of reduced dimensionality HN(C)N experiments presented here is that they provide rapid identification of sequential HSQC peaks through $F_2$($^{15}$N)–$F_3$($^1$H) planes of their respective spectra along the co-evolved $F_1$(NC) dimension. The process has been illustrated experimentally in **Fig. 4**. As evident, the sequential ($i$+1) HSQC peaks can be identified on the $F_2(^{15}N)$–$F_3(^{1}H)$ planes of these spectra at $F_1= N^{+}_{i+1} / N^{-}_{i+1}$ chemical shifts identified for residue, *i*. Compared to the routine experiments (like HNCA, HN(CA)CO, CBCANH, etc), which also provide the identical feature along the $^{13}C$ dimension, the advantage here is the presence of opposite peak signs for self and sequential correlations; therefore the sequential HSQC peaks ($H_{i+1}$, $N_{i+1}$) can be easily differentiated from the self HSQC peaks ($H_i$,$N_i$). The process is facilitated further (both in terms of accuracy and speed) by the random and well dispersed nature of $^{15}N^{+/-}$ chemical shifts present along the jointly sampled $F_1$(NC) dimension. Further, a particular advantage is that there are two different sets of sequential correlations i.e. $^{15}N^+$ (up-field domain) and $^{15}N^-$ (down-field domain). Thus, any ambiguity in the identification of (*i* → *i*+1) sequential correlation using up-field set of peaks can be resolved using down-field set of peaks and vice versa.

A point to be mentioned here is that the separation –between sum and difference frequencies (i.e. $^{15}N^+$ and $^{15}N^-$) in the final spectra of (4,3)D-h<u>NCO</u>caNH and (4,3)D-h<u>N</u>co<u>CA</u>NH experiments– has been achieved by keeping the offsets of $^{13}C$ channel at 70 and 188 ppm, respectively. In order to achieve this separation between the addition and subtraction frequencies (i.e. $^{15}N^+$ and $^{15}N^-$), one should be careful about (i) the spectral width used along the co-evolved dimension and (ii) the offset frequencies used along $^{13}C$ channel. The combined analysis of average $^{13}C^{\alpha}$ and $^{13}C'$ chemical shifts of all the amino acids (**Fig. 1**; data taken from BMRB) and the **Eqns 1 and 2** revealed that for a given protein the two frequency regions (i.e. $^{15}N^+$ and $^{15}N^-$) will be separated from each other (i) if the $^{13}C$ carrier is kept 1–2 ppm away from the most down fielded shifted $^{13}C$ chemical shift; this is typically close to ~188 and ~70 ppm, respectively, for $^{13}C'$ channel [in case of (4,3)D-h<u>NCO</u>caNH] and for $^{13}C^{\alpha}$ channel [in case of (4,3)D-h<u>N</u>co<u>CA</u>NH] and (ii) the spectral width along the co-evolved $F_1$(NC) dimension is increased accordingly; typically close to ~80 and ~150 ppm, respectively, for (4,3)D-h<u>NCO</u>caNH and (4,3)D-h<u>N</u>co<u>CA</u>NH spectra. Frequency selection along all the indirect dimensions has been done by States-TPPI method [30] (in the coupled N-C dimension, quadrature detection is performed only for the $^{15}N$) and the data is acquired in a way that it is processed using normal Fourier Transformation (for detail see this reference [31]).

Overall, the true sequential HSQC peak ($H_{i+1}$,$N_{i+1}$) will be of opposite sign on both the $^{1}H$–$^{15}N$ planes of spectrum at $F_1= N^{+}_{i+1} / N^{-}_{i+1}$ chemical shifts identified for residue, *i*. An illustrative example has been shown in **Figure 3** depicting the use of correlations observed in the $F_2(^{15}N)$–$F_3(^{1}H)$ planes of (4,3)D-h<u>N</u>co<u>CA</u>NH spectrum of 8M urea-denatured UNC60B (for residues Thr36 to Val39) at the $^{15}N^{+}_{i+1} / ^{15}N^{-}_{i+1}$ chemical shifts identified for residue *i* for establishing a sequential (*i* → *i*+1) connectivity between the HSQC peaks. As evident, the spectrum provides rapid and unambiguous identification of sequential HSQC peaks which is particularly important while performing sequential assignment of intrinsically/partially unstructured proteins, proteins containing repetitive amino acids (e.g....TSAAGTTTE...) or repetitive stretches of amino acids (e.g. ...QPLAGA... QPLAGA...) as well as for protein folding/unfolding studies by NMR. Backbone assignment following conventional strategies has always been a great challenge in all such cases where the poor dispersion of $^{13}C^{\alpha/\beta}$ chemical shifts (e.g. as in case of CBCANH) and their dependence on the amino-acid type result in the crowding of self and sequential HSQC peaks on $^1H-^{15}N$ planes of these spectra at the degenerate carbon chemical shifts. Likewise in HN(C)N spectrum, the $^{15}N$ chemical shift degeneracy may also lead to ambiguities in the identification of sequentially connected HSQC peaks (more details are given below). Overall, the reduced dimensionality HN(C)N experiments presented here would greatly facilitate the resonance assignment of proteins exhibiting crowded HSQC spectra (i) by resolving the problems arising because of amide $^{15}N$ shift degeneracy and (ii) by providing rapid and unambiguous identification of sequential HSQC peaks.

## *Assignment Protocol:*

The assignment protocol based on the reduced dimensionality HN(C)N experiment [i.e. (4,3)D-h<u>NCO</u>ca<u>N</u>H or (4,3)D-h<u>N</u>co<u>CA</u>NH] has been illustrated schematically in **Figure 4** using an example stretch of amino acids -XYZ- where X represents residue *i*. As illustrated in **Fig. 4**, a sequential connectivity (*i* to *i*+1) between the HSQC peaks can be established following the routine assignment procedure by matching the self (here $N_i^+$ and $N_i^-$) and sequential (here $N_{i+1}^+$ and $N_{i+1}^-$) correlation peaks identified along the $F_1$(NC) dimension of $F_1$(NC)–$F_3$($^1$H) planes at $^{15}N_i$ chemical shift. This is experimentally demonstrated in **Fig. 3A** using the $F_1$-$F_3$ strips from (4,3)D-h<u>N</u>co<u>CA</u>NH spectrum of unfolded UNC60B. Like HN(C)N, the remarkable feature of these RD HN(C)N spectra is that they provide easy discrimination between self and sequential correlation peaks because of their opposite peak signs and thus no other complementary experiment is required for discriminating them. The other advantage of the protocol is that it does not involve cumbersome search through various $F_2$ planes of the 3D spectrum for matching the frequencies along the $F_1$ dimension. Rather, a sequential *i* to *i*+1 connectivity between two HSQC peaks can be established unambiguously on the $F_2$($^{15}$N)–$F_3$($^1$H) planes of spectrum at $F_1 = N_{i+1}^+$ or $F_1 = N_{i+1}^-$ **(Fig. 4B)**. The process has been demonstrated experimentally in **Fig. 3** using $F_2$($^{15}$N)–$F_3$($^1$H) planes of (4,3)D-hNco<u>CA</u>NH spectrum of 8M urea denatured UNC60B for residues Thr36-Val39. After establishing all the possible *i* → i+1 connectivities, the stretches of sequentially connected HSQC peaks are then mapped on to the primary sequence for assigning them sequence specifically. For this, the strategy makes use of triplet specific peak patterns present in these spectra as described in **Appendix III** and **Figure S2** (Supplementary material). As evident from the figure, these are the glycines and prolines which are responsible for special patterns of self and sequential correlation peaks. These special patterns provide identification of certain specific triplet sequences and thus serve as check points for mapping the stretches of sequentially connected HSQC peaks on to the primary sequence for final assignment. The process has been illustrated schematically in **Fig. 4C**. A point to be mentioned here is that like the basic HN(C)N, these experiments can also be modified to generate additional amino-acid specific (i.e. alanines and serines/threonines specific) patterns of self and sequential correlation peaks and hence enable identification of additional triplets of residues [28,29]. This becomes important in those cases when the protein sequence has only a few glycines and prolines and they are far removed along the sequence. As described previously [28,29], the additional check points can be generated simply by changing the bandwidth of the 180° inversion pulse on $^{13}C^{\alpha}$ channel during the $\tau_{CN}$ evolution period (red encircled pulse, **Fig. 2**). Accordingly the experiment can have three variants: (a) glycine variant (described here in the present paper, (b) alanine variant (which provides identification of triplet stretches containing glycines/alanines and prolines), and (c) serine/threonine variant (which provides identification of triplet stretches containing serines/threonines and prolines). However, to avoid the confusion, the experimental results and the assignment protocol based on the normal experiment have been presented here.

Following the above protocol, the complete sequence specific assignment of backbone ($^1H$, $^{15}N$, $^{15}N^+$, and $^{15}N^-$) resonances is established initially and then the backbone $^{13}C^{\alpha}$ or $^{13}C'$ chemical shifts are determined as:

$$^{13}C_i^{obs} = (^{15}N_{i+1}^+ - ^{15}N_{i+1}^-)/2\kappa + ^{13}C_{offset} \qquad (3)$$

where '$k$' is the scaling factor and is equal to $\gamma(^{13}C)/\gamma(^{15}N)=2.48$. Thus, the single reduced dimensionality HN(C)N experiment provides direction specific sequential assignment walk without involving any complementary experiment. Even the explicit side chain assignment would not be very necessary to decide on the correctness of the sequential assignment because of the large number of various check points that are generally available. The assignment protocol is thus very simple, swift and well suited for automation as well. Indeed, we are in a process to automate the protocol as AUTOBA+ (the updated version of AUTOBA [15]) which would also be able to handle the orthogonal projection planes of these RD HN(C)N experiments (as shown in **Fig S5**) for the assignment of well-behaved medium sized (MW ~ 12-15 kDa) folded proteins.

The described sequential assignment walk protocol relies mainly on the presence of sufficient number of check points and moreover these should be well-dispersed all along the sequence. These check points are generally obtained around either glycines or alanines/glycines, or serines/threonines (depending upon the variant of the experiment used). However in case when these residues are far removed along the sequence of polypeptide chain and/or when there are several ambiguous breaks in sequential connectivities due to missing amide cross-peaks (mainly due to conformational line broadening effect), the above sequential assignment walk protocol may fail. Further, in case when a protein exhibits degeneracy in both amide $^1$H and $^{15}$N shifts (i.e. when there is overlap of HSQC peaks), the strategy may lead to ambiguities in backbone assignment. In such a case, residue type identifications will help to remove the ambiguity. This is exactly the same deadlock situation encountered here for the assignment of unfolded UNC60B protein (152 residues; MW~17 kDa). **Figure S6A** (Supporting Information) depicts the $^1$H-$^{15}$N HSQC spectrum of the unfolded UNC60B denatured in 8 M Urea at pH 6.0 and 298K. The limited chemical shift dispersion of the backbone amide $^1$H resonances (within the range of ~0.75 ppm) indicated that the polypeptide chain of UNC60B is highly denatured under the experimental conditions. However, due to extensive conformational exchange and peak overlap at 298 K, only 88 amide-cross peaks were discerned above the noise level out of 148 expected non-proline peaks. For these 88 peaks, the sequential connectivities were established for 79 (~ 90 %) peaks unambiguously (like as depicted in **Fig. S6B and S6C)**. However, because of lack of sufficient number check points, the transformation of stretches of sequentially connected HSQC peaks into the final sequence specific assignment was not possible. In such a situation, the sequential $^{13}C^{\alpha}$ and $^{13}C^{\beta}$ chemical shift information derived from 3D-CBCAcoNH experiment has been used to transform these stretches into the final assignment. Therefore, the unambiguous sequential walk approach based on these reduced dimensionality tailored HN(C)N experiments combined with a 3D CBCAcoNH experiment (for amino acid type identification) provides an alternatively robust and efficient approach for sequential backbone resonance assignment (however, the actual combination of experiments will depend upon the situations at hand). Following this approach, we were able to assign 90% of backbone ($^1$H, $^{15}$N, and $13C^{\alpha/\beta}$) chemical shifts of unfolded UNC60B sequence specifically **(Fig. S6, Supporting Information)**. A summary of all the observed sequential connectivities in the (4,3)D-hNcoCANH spectrum of unfolded UNC60B protein has been shown by underlining along the sequence **(Fig. S6D)**. As evident from the assignment of unfolded UNC60B, only the N- and C-terminal residues of the polypeptide chain give rise to observable peaks in the HSQC spectrum indicating extensive line broadening of amide cross peaks signals from the core of the polypeptide chain.

Finally, the guidelines -regarding in what situation which of the two experiments should be used- are also required to be mentioned here. As evident from **Fig S1** and the experimental results depicted here, the (4,3)D-hNcoCANH experiment provides relatively a better dispersion of chemical shifts compared to (4,3)D-hNCOcaNH experiment, therefore the former should be used preferably. However for proteins containing repetitive amino acids (like …TSAAGTTTE…) or repetitive stretches of amino acids (like …GTSDE…GTSDE…), the (4,3)D-hNCOcaNH experiment will be a better option compared to (4,3)D-hNcoCANH because of the random nature of backbone $^{13}$C' chemical shifts.

### *Advantage over the Basic Experiment in its Application to Unfolded Proteins:*

Unfolded proteins (including intrinsically unstructured, partially disordered and proteins denatured for studying their folding mechanisms) have always posed a great challenge for NMR investigations because of poor chemical shift dispersion of amide $^1$H and carbon resonances. Pulse sequences relying heavily on well dispersed nitrogen chemical shifts –like HN(C)N [10,13]- have proven extremely useful in this context. They provide direct sequential amide $^1$H$^N$ and $^{15}$N correlations along the polypeptide chain and exploit the $^{15}$N chemical shift dispersion along two of the three dimensions, which is generally very good for unfolded proteins. However, the HN(C)N based assignment protocols may fail when the HSQC peaks of sequential residues have degenerate amide $^{15}$N chemical shifts. The fact has been demonstrated here in **Figure 5** (schematically) **and Figure 6** (experimentally) using HN(C)N as an example. The two situations may arise as discussed below:

(i) When two HSQC peaks from self (*i*) and sequential (*i*+1) residues have almost the same $^{15}$N chemical shift (Case-I, **Figure 5**): In this case, the sequential correlation peak (i.e. H$_i$N$_{i+1}$) in the $F_1$($^{15}$N)-$F_3$($^1$H) plane of the HN(C)N spectrum can disappear because of opposite sign and higher intensity of self-correlation peaks (i.e. H$_i$N$_i$) in this spectrum as shown in **Fig. 5B**, and sequential assignment walk may end up with an ambiguity that means (a) residue *i* is present at the C-terminal or (b) it is followed by a proline or (c) the sequential (*i*+1) residue has the degenerate $^{15}$N chemical shift (i.e. $^{15}$N$_i$ = $^{15}$N$_{i+1}$). However, the ambiguity is resolved here in the reduced dimensionality HN(C)N experiments by breaking the $^{15}$N degeneracy condition by using linear combination of $^{15}$N and $^{13}$C$^{\alpha}$/$^{13}$C' chemical shifts. For example, the $F_1$(NC)-$F_3$($^1$H) strip of RD-HN(C)N spectrum in **Fig. 5C** clearly reveals that residue is not followed by proline; instead self (*i*) and sequential (*i*+1) HSQC peaks have the degenerate $^{15}$N chemical shifts.

(ii) When the residues (here *i* and *j*) having the degenerate $^{15}$N chemical shifts in the HSQC spectrum (i.e. $^{15}$N$_i$ = $^{15}$N$_j$) are sequentially connected to residues with almost same $^{15}$N chemical shift (i.e. $^{15}$N$_{i+1}$ = $^{15}$N$_{j+1}$) Case-II, **Figure 5**): In this case, the HN(C)N spectrum (or the $F_2$($^{15}$N)-$F_3$($^1$H) plane at $F_1$ = $^{15}$N$_{i+1}$ = $^{15}$N$_{j+1}$; as shown in **Fig. 5E**) may fail to identify the sequential (*i*+1 or *j*+1) HSQC peaks corresponding to residues *i* and *j*, unambiguously. Under such situation, one has to consider both the possibilities as shown in **Fig. 5E** which elaborates the analysis. However, in reduced dimensionality HN(C)N experiments such an ambiguity (arising because of degenerate $^{15}$N chemical shift) can be resolved along $F_1$(NC) dimension (which is devoid of the $^{15}$N degeneracy condition), where now $i \rightarrow i+1$ and $j \rightarrow j+1$ sequential amide correlations will appear on different $F_2$($^{15}$N)-$F_3$($^1$H) planes i.e. at $^{15}N_{i+1}^{+}/^{15}N_{i+1}^{-}$ and $^{15}N_{j+1}^{+}/^{15}N_{j+1}^{-}$, respectively, as clear from **Figure 5**.

Therefore in terms of data analysis particularly when two neighboring amino acid residues have nearly identical backbone $^{15}$N chemical shifts, the presented reduced dimensionality tailored HN(C)N experiments perform relatively better compared to the basic HN(C)N experiment (a situation generally seen for unfolded or intrinsically unstructured polypeptide chains). To highlight the relative ease and speed of sequential assignment of any protein based on these new experiments, the representative 2D $F_1$-$F_3$ strip plots and 2D $F_2$-$F_3$ planes of these spectra have been compared to those of basic HN(C)N spectrum in **Fig. 6.** Depending upon the situation in hand, **Fig. 6** shows comparison of the spectral features for establishing a sequential *i* to *i*+1 connectivity between Leu17 and Leu18 of unfolded UNC60B. As evident from **Fig. 6B** and **6C**, the identification of sequential (*i*+1) correlation peak corresponding to HSQC peak (H$_i$N$_i$; here it is Leu17) is ambiguous on $F_2$-$F_3$ planes of HN(C)N spectrum at $F_2$=N$_i$, and one has to check for all the cross peaks corresponding to N$_{i+1}$ chemical shifts (identified initially from the $F_1$-$F_3$ plane of HN(C)N spectrum at $F_2$=N$_i$). In **Fig. 6C,** these possibilities are shown by the blue horizontal lines. However, on the other hand, a sequential *i* to *i*+1 connectivity between the two HSQC peaks can be established in a very simple, swift and accurate manner following the unambiguous RD HN(C)N based approach i.e. a true sequential HSQC peaks ($H_{i+1}$,$N_{i+1}$) will be present on both the $F_2$-$F_3$ planes of these spectra at $F_1 = N_{i+1}^+/N_{i+1}^-$ (identified from the $F_1$-$F_3$ plane of RD HN(C)N spectrum at $F_2$=N$_i$, **Fig. 6D**) These are shown by the grey horizontal lines connecting the two red peaks in **Fig. 6E.**

**Concluding Remarks:**

In conclusion, two novel reduced dimensionality experiments -(4,3)D h<u>N</u>co<u>CA</u>NH and (4,3)D h<u>NCO</u>caNH- and an efficient assignment protocol has been demonstrated here for obtaining un-ambiguous and accurate assignment of backbone amide ($^1$H and $^{15}$N) and $^{13}$C$^\alpha$ or $^{13}$C' resonances. The protocol is basically an improvisation of the previously reported HN(C)N based protocol [14] and is based on the fact that linear combination of backbone $^{15}$N and $^{13}$C (here either $^{13}$C$^\alpha$ or $^{13}$C') chemical shifts offers relatively better dispersion and randomness compared to individual chemical shifts **(Fig. S1)**; thus helps to resolve the ambiguities arising because of amide $^{15}$N chemical shift degeneracy (as in case of HN(C)N based assignment protocol). Overall, the experiments would greatly facilitate the assignment of crowded HSQC spectra especially of intrinsically/partially unstructured proteins and medium sized helical proteins which in general exhibit high degree of amide shift degeneracy. The performance of the experiments and the assignment protocol has been demonstrated here two folded proteins (human ubiquitin and bovine apo calbindin-d9k) and on a highly denatured UNC60B protein unfolded in 8M urea.

Compared to currently available strategies, the strength of the protocol lies in the fact that the single RD HN(C)N experiment provides the complete sequence specific assignment of backbone amide resonances. The method is very well suited for automated data analysis as well where keeping the total number of spectra small is highly desirable in order to avoid the complications arising because of inter-spectral variations of chemical shifts; the problem is common in proteins which are unstable in solution or tend to precipitate in matter of days. The method will also serve as a valuable NMR assignment tool in drug discovery research programs especially for SAR by NMR (i.e. studying Structure Activity Relations by NMR and is the most commonly used method to verify binding, stoichiometry, and identification of the binding site on the protein targets [32,33]). Likewise, the method can also be used to re-establish the lost resonance assignment of proteins upon ligand binding or upon a mutation. Taken together, the experiments presented here will serve as an efficient backbone amide assignment tool for various protein NMR studies.

**Acknowledgement:**

This work is being financially supported by the Department of Science and Technology under SERC Fast Track Scheme (Registration Number: **SR/FT/LS-114/2011**) for carrying out the research work. I would also like to acknowledge the High Field NMR Facility at Centre of Biomedical Magnetic Resonance, Lucknow, India. For demonstrating the method on an unfolded protein, the $^{13}$C/$^{15}$N labeled sample of protein UNC60-B was gifted by Dr. Ahish Arora of Molecular Structural Biology Division, CDRI, Lucknow. I wish to dedicate this work to Prof. Ramakrishna V Hosur, TIFR, Mumbai, India who has always supported and encouraged me to pursue research in NMR methodology development.

## Figures:

**Figure 1:** Schematic illustrations of the selected coherence transfer pathways employed in reduced dimensionality experiments (4,3)D- hNCOcaNH **(A)** and (4,3)D- hNcoCANH **(B)**, respectively. The magnetization flow from H$^N$(*i*) is shown and the frequency labeling of the appropriate nuclei are indicated where, $2T_N$, $2\tau_C$, $2\tau_{CN}$, and $2\tau_N$ are the delays during which the transfers indicated by the arrows take place in the pulse sequence. **(C)** and **(D)** represent the pulse sequences for RD (4,3)D hNCOcaNH and (4,3)D hNcoCANH experiments, respectively. Narrow (hollow) and wide (filled black) rectangular bars represent non-selective 90° and 180° pulse, respectively. Unless indicated, the pulses are applied with phase $x$. Proton decoupling using the DIPSI-2 decoupling sequence [34] with field strength of 6.3 kHz is applied during most of the $t_1$ ($^{15}$N/$^{13}$C) and $t_2$ ($^{15}$N) evolution periods, and $^{15}$N decoupling using the GARP-1 sequence [35] with the field strength 0.9 kHz is applied during acquisition. Standard Gaussian cascade pulses [23] –shape Q3 for 180° inversion/refocusing (filled black and grey; width 200 μs) and shape Q5 for 90° excitation (hollow, width 310 μs)– were used along the carbon channel. The bandwidth of the $^{13}$C$^\alpha$ pulses (either for excitation, inversion or refocusing) is adjusted so that they cause minimal excitation of carbonyl carbons and that of 180° $^{13}$C' shaped pulse so that they cause minimal excitation of $^{13}$C$^\alpha$. The blue encircled red pulse on $^{13}$C$^\alpha$ channel is crucial for tuning the experiment for generation of different check points in the final spectrum (like glycines/alanines and serines/threonines, for detail see these references [28,29]). Yellow pulses were applied for compensation of off-resonance effects (Bloch-Siegert phase shift) [36]. The values for the individual periods containing $t_1$ evolution of $^{15}$N nuclei are: A = $t_1/2$, B = $T_N$, and C = $T_N - t_1/2$. In (4,3)D hNCOcaNH pulse sequence, the $^{13}$C' nuclei are co-evolved with $^{15}$N nuclei in a semi-constant time manner, and the values for the individual periods containing $t_1$ evolution of $^{13}$C' nuclei are: $\tau_C^a = t_1/2$, $\tau_C^b = \tau_C$ and $\tau_C^c = \tau_C - t_1/2$. In (4,3)D hNcoCANH pulse sequence, the $^{13}$C$^\alpha$ nuclei are co-evolved with $^{15}$N nuclei in constant time manner and the values for the individual periods containing $t_1$ evolution of $^{13}$C$^\alpha$ nuclei are: $G = t_1/2$, $H = \tau_{CN} - \tau_C$, and $J = \tau_{CN} - t_1/2$. The values for the individual periods containing $t_2$ evolution of $^{15}$N nuclei are: D = $\tau_N - t_2/2$, E = $\tau_N$, and F = $t_2/2$. The other delays are set to λ = 2.5 ms, $\kappa$ = 5.4 ms, δ = 2.5 ms, $\tau_C$ = 4.5 ms, $T_N$ = 14.0 ms, $\tau_{CN}$ = 12.5 ms and $\tau_N$ = 13.5 ms. The $\tau_{CN}$ must be optimized and is around 12-15 ms. The phase cycling for the experiment is Φ$_1$ = 2($x$), 2(-$x$); Φ$_2$ = Φ$_3$= $x$, -$x$; Φ$_4$ = Φ$_5$ = $x$; Φ$_6$ = 4($x$), 4(-$x$); and Φ$_{receiver}$ = 2($x$), 4(-$x$), 2($x$). The frequency discrimination in $t_1$ and $t_2$ has been achieved using States-TPPI phase cycling [30] of Φ$_1$ and Φ$_5,$ respectively, along with the receiver phase. The gradient (sine-bell shaped; 1 ms) levels are as follows: G$_1$=30%, G$_2$=30%, G$_3$=30%, G$_4$=30%, G$_5$=50%,

$G_6$=80% and $G_7$=20% of the maximum strength 53 G/cm in the z-direction. The recovery time after each gradient pulse was 160 μs. Before detection, WATERGATE sequence [24] has been employed for better water suppression.

Kumar D22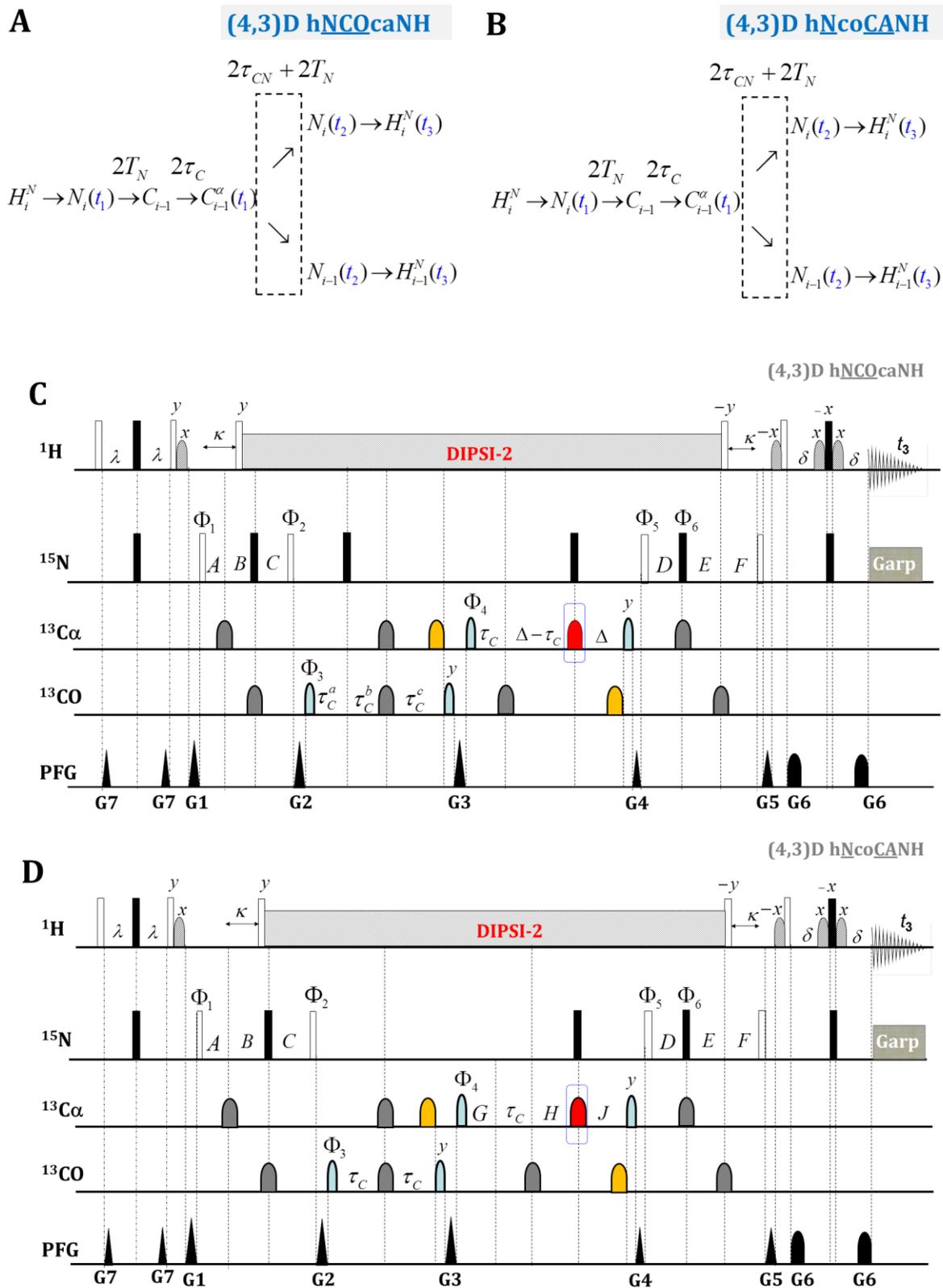
Kumar D

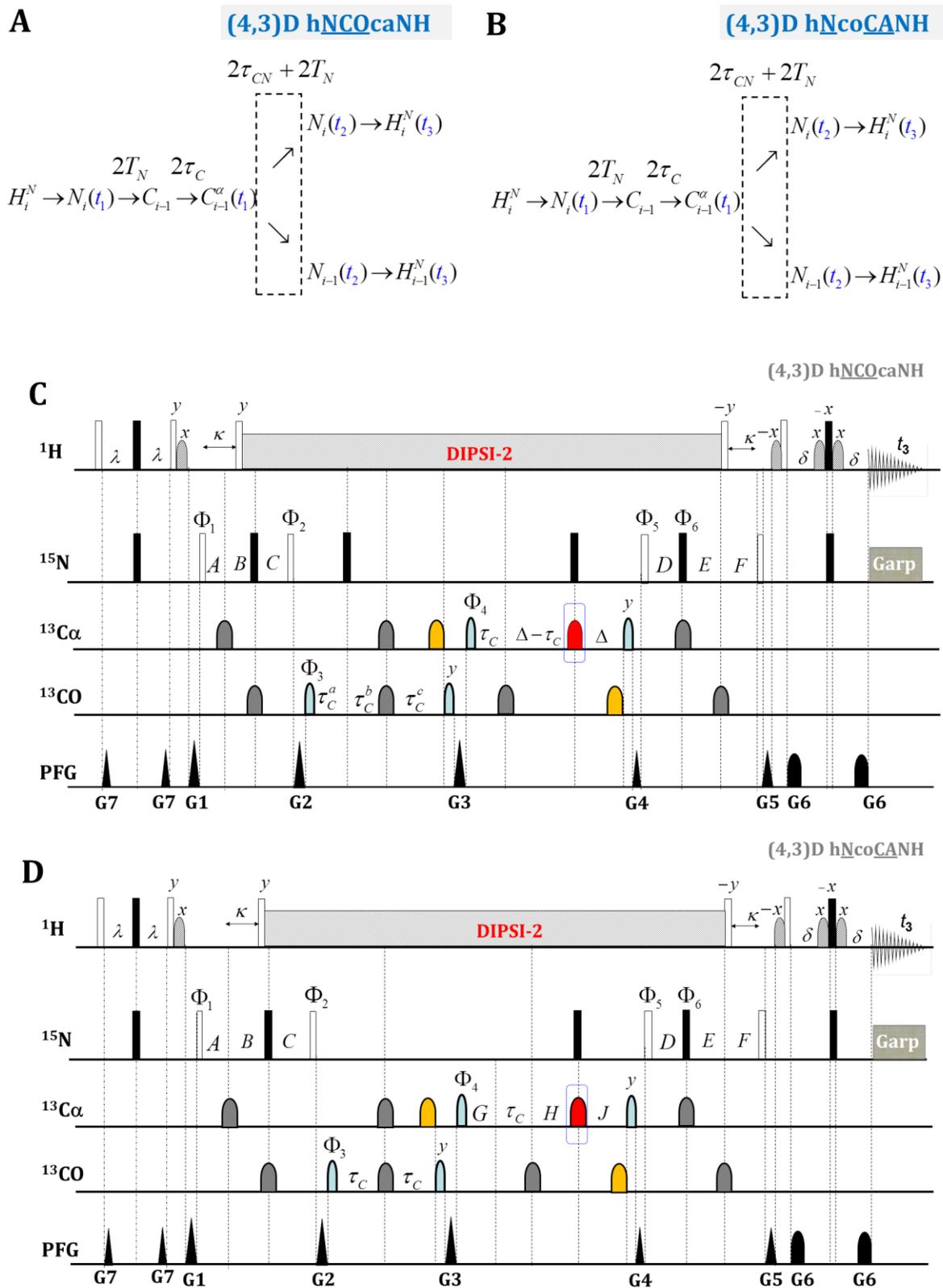

**Figure 2:** Schematic representation of the three-dimensional (4,3)D-hNCOcaNH/(4,3)D-hNcoCANH spectrum. The correlations observed in the $F_1(NC)$–$F_3(^1H)$ plane at the $^{15}N$ chemical shift of residue *i* are shown on left side and the correlations observed in the $F_2(^{15}N)$–$F_3(^1H)$ planes at $F_2 = N_i^+ / N_i^-$ and $F_2 = N_{i+1}^+ / N_{i+1}^-$ are shown on the right side. Squares and circles represent the self and sequential peaks, respectively. Red and black represent positive and negative phase of peaks, respectively.

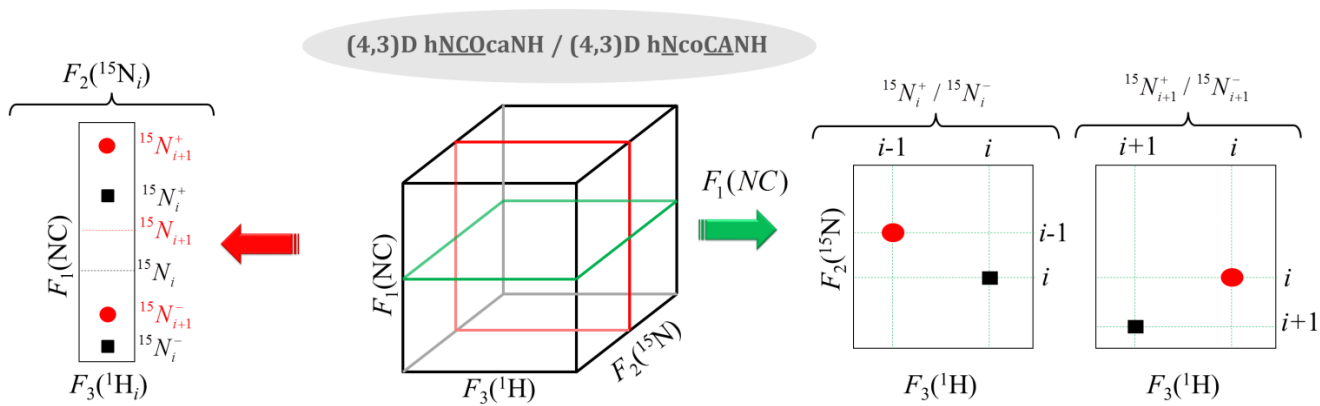

**Figure 3:** An illustrative example showing use of correlations observed in the $F_2(^{15}N)$–$F_3(^1H)$ planes (of (4,3)D-h<u>N</u>co<u>CA</u>NH spectrum of 8M urea-denatured UNC60B) at the $^{15}N^+_{i+1}$ / $^{15}N^-_{i+1}$ chemical shifts identified for residue $i$ for establishing a sequential ($i \rightarrow i+1$) connectivity between the backbone amide correlation peaks of $^1H$-$^{15}N$ HSQC spectrum (like in case of HN(C)N based assignment protocol [14]). Red and black contours represent positive and negative phase of the peaks, respectively. A sequential amide cross peak ($H_{i+1}$, $N_{i+1}$) exists in both the $F_2(^{15}N)$–$F_3(^1H)$ planes of the spectrum at the $^{15}N^+_{i+1}$ and $^{15}N^-_{i+1}$ chemical shifts identified for the residue $i$. The other advantage is the opposite peak signs for the self ($H_i$, $N_i$) and sequential ($H_{i+1}$, $N_{i+1}$) amide correlation peaks. These spectral features help to reduce the search for the sequential amide correlations rather than making the search in various planes of the 3D spectrum (as is the case with other presently used assignment strategies).

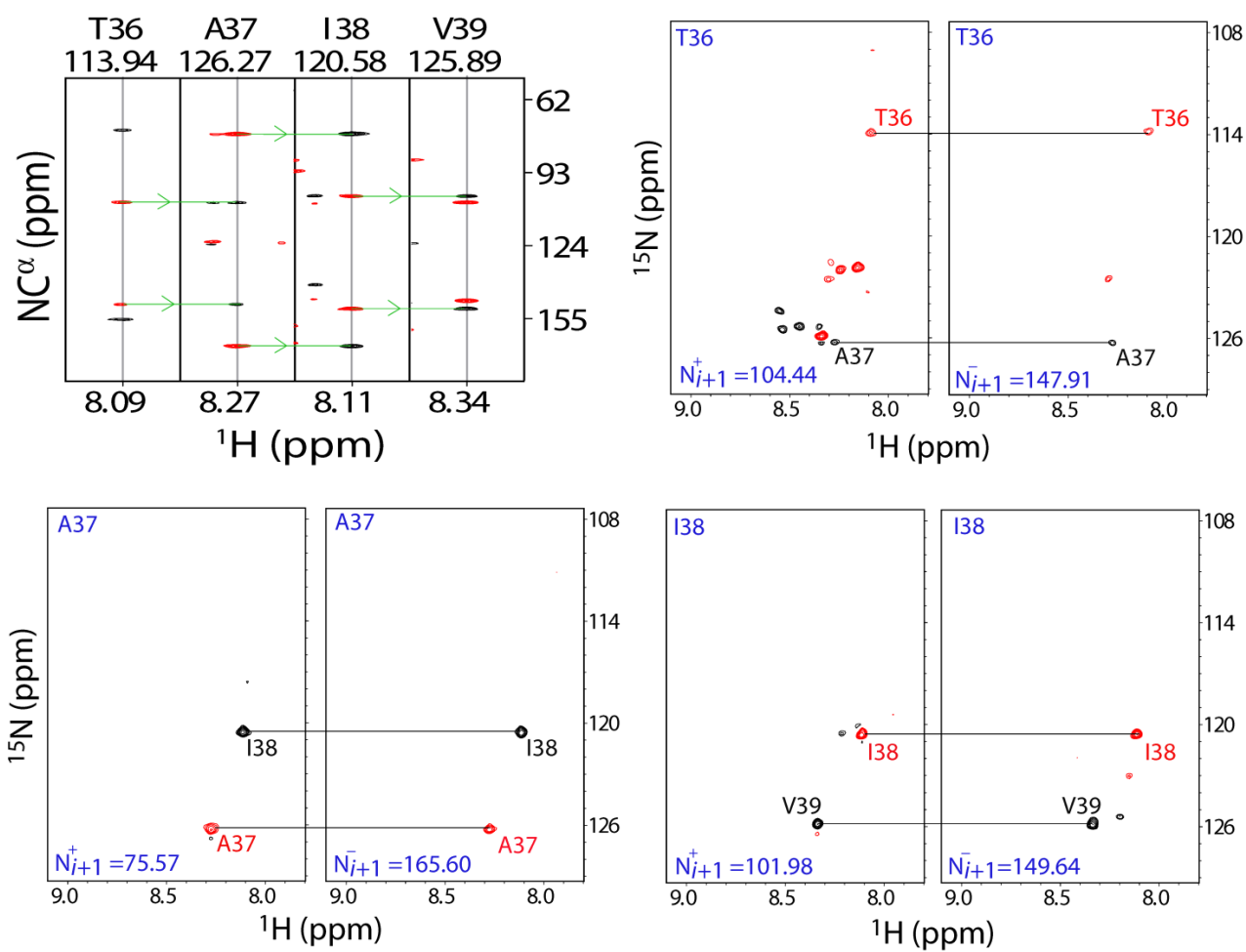

**Figure 4:** Schematic showing sequential assignment walk protocol based on the reduced dimensionality HN(C)N experiments. In **(A)** and **(B)**, the strategy of establishing the sequential (*i* to *i*+1) connectivity has been described using the polypeptide sequence –XYZ–. In **(C)**, the sequential assignment walk procedure –through the $F_1$(NC)–$F_3$($^1$H) planes of the reduced dimensionality HN(C)N spectra– using the triplet specific peak patterns has been shown. An arbitrary amino acid sequence is chosen to illustrate the start, continue, check, and break points during the sequential assignment walk. Squares and circles represent the self (intra-residue) and sequential (inter-residue) peaks, respectively. Red and black represent positive and negative phase of peaks, respectively. The patterns of positive and negative peaks are drawn according to the triplet specific peak patterns contained in this spectrum (as shown in **Fig. 4**). The residue identified on the top of each strip identifies the central residue of the triplet.

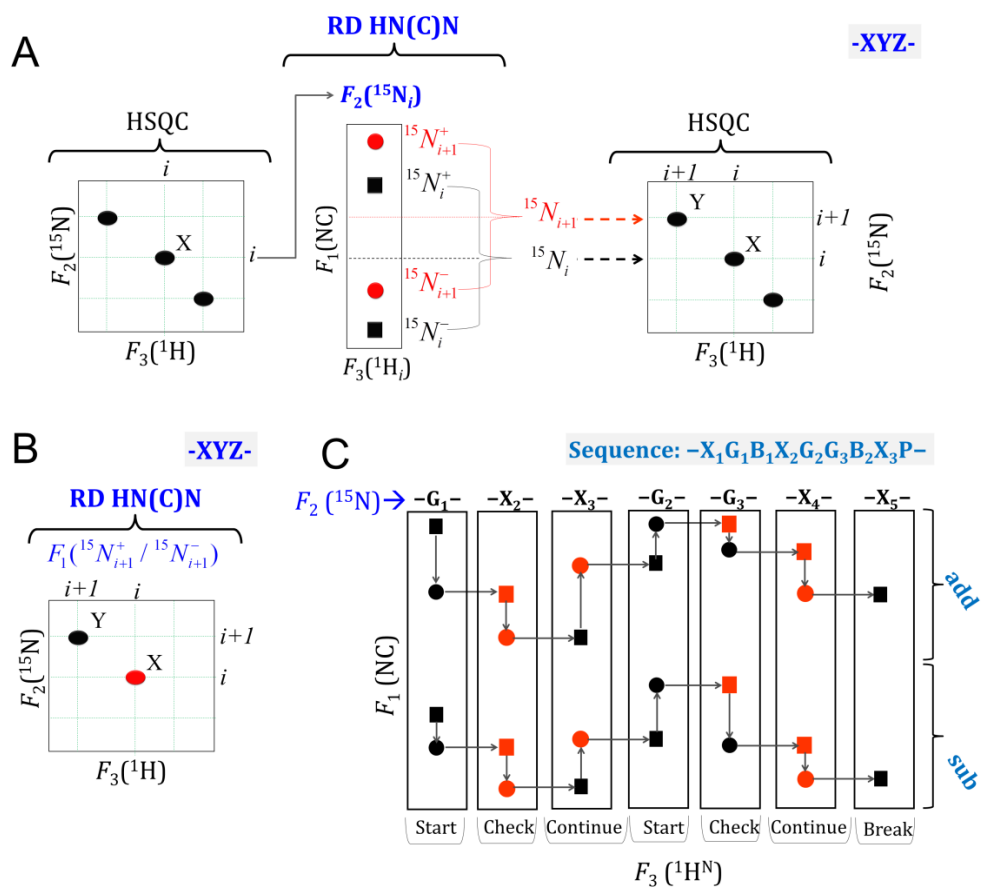

**Figure 5:** Schematic showing the advantage of reduced dimensionality HN(C)N experiment for resolving the ambiguities arising because of degenerate 15N chemical shifts as in case of HN(C)N based assignment protocol.

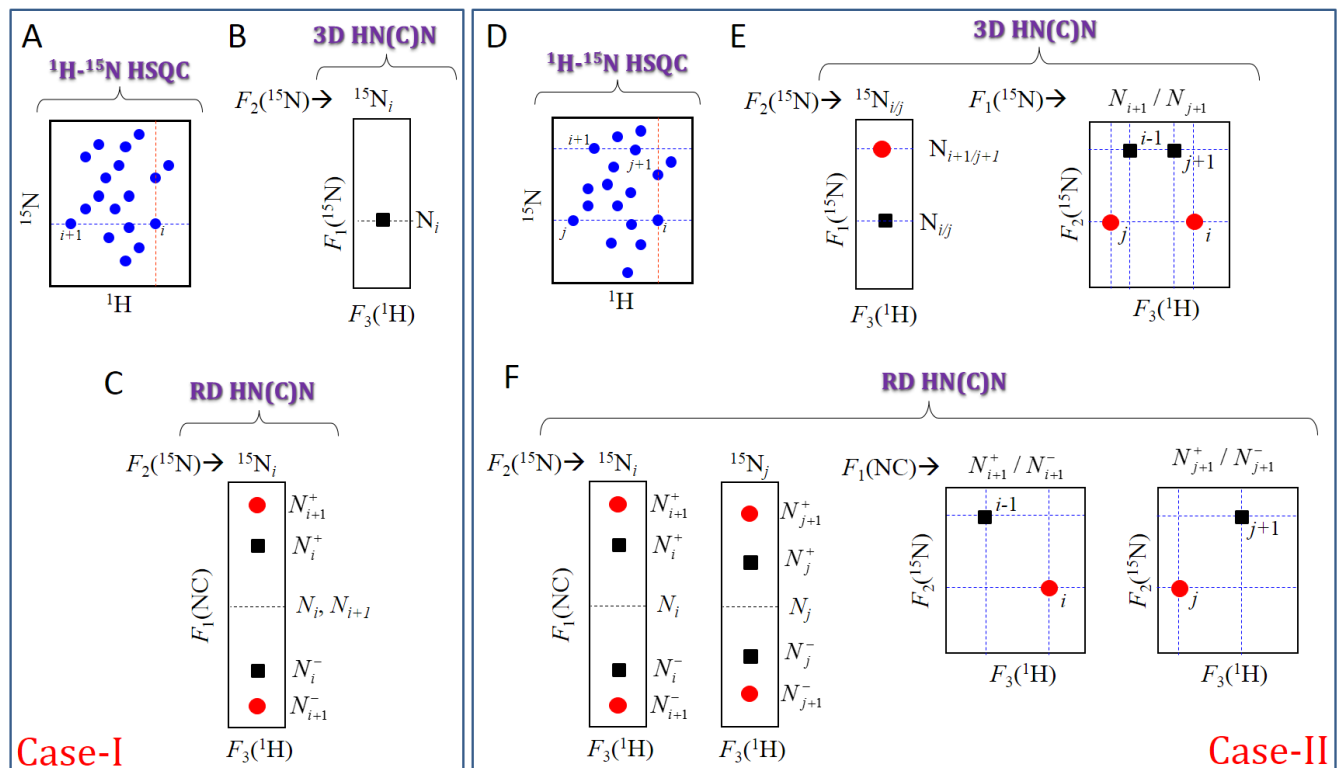

Where RD HN(C)N represents either (4,3)D h<u>NCO</u>caNH *Or* (4,3)D h<u>N</u>co<u>CA</u>NH

**Figure 6:** Experimental demonstration on unfolded UNC60B protein, of the advantage of RD (4,3)D-HN(C)N based assignment protocol for resolving the ambiguities arising due to degenerate $^{15}$N chemical shifts. **(A)** A part of $^1$H-$^{15}$N HSQC spectrum of unfolded UNC60B (in 8 M Urea at pH 6.0 and 298 K) showing relative positions of sequentially connected HSQC peaks: Asp16 to His19. **(B)** $F_1(^{15}$N$)$-$F_3(^1$H$)$ strips and **(C)** $F_2(^{15}$N$)$-$F_3(^1$H$)$ planes of 3D-HN(C)N spectra showing ambiguity in finding the sequential correlation to spin system Leu17. **(D)** $F_1(NC^\alpha)$-$F_3(^1$H$)$ strips and **(E)** $F_2(^{15}$N$)$-$F_3(^1$H$)$ planes of of (4,3)D-h<u>N</u>co<u>CA</u>NH spectrum showing unambiguous sequential connection between spin systems Leu17 and Leu18. The $F_2$ ($^{15}$N) values are shown at the top for each $F_1$-$F_3$ strip. In **(D),** the top and bottom halves belong to the 'minus' and 'plus' domains of the spectrum, respectively.

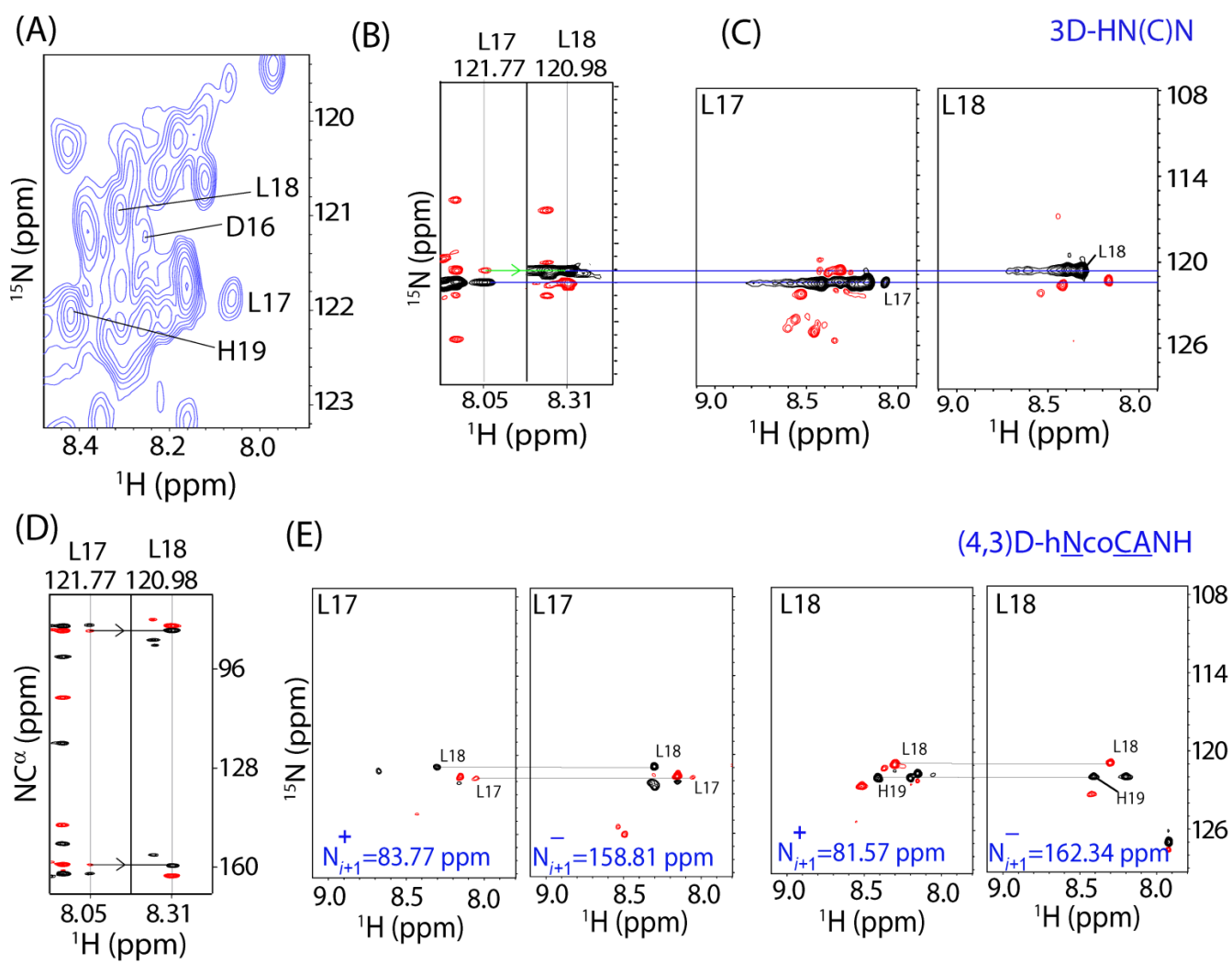

# Supplementary Material:

# Reduced Dimensionality tailored HN(C)N Pulse Sequences for Efficient Backbone Resonance Assignment of Proteins through Rapid Identification of Sequential HSQC peaks

Dinesh Kumar

Centre of Biomedical Magnetic Resonance, SGPGIMS Campus, Raibareli Road, Lucknow-226014, India

**\*Address for Correspondence:**

Dr. Dinesh Kumar
(Assistant Professor)
Centre of Biomedical Magnetic Resonance (CBMR),
Sanjay Gandhi Post-Graduate Institute of Medical Sciences Campus,
Raibareli Road, Lucknow-226014
Uttar Pradesh-226014, India
Mobile: +91-9044951791,+91-8953261506
Fax: +91-522-2668215
Email: dineshcbmr@gmail.com
Webpage: http://www.cbmr.res.in/dinesh.html

# Appendix *I*

The particular advantage of new experiments is the dispersion of peaks in the sum and difference frequency regions which can be different and this facilitates analysis of the spectrum by simultaneously connecting ($^{15}$N+$^{13}$C) and ($^{15}$N+$^{13}$C) correlations (here $^{13}$C is either backbone $^{13}$C' or $^{13}$C$^\alpha$). The advantage is elicited from the fact that the linear combination of $^{15}$N$_i$ and $^{13}$C$^\alpha_{i-1}$/$^{13}$C'$_{i-1}$ chemical shifts provides dispersion better compared to the dispersion of the individual chemical shifts. The fact has been demonstrated here in **Fig. S1** using average $^{15}$N, $^{13}$C$^\alpha$ and $^{13}$C' chemical shifts derived from the BMRB database [1] for all the 20 amino acids. As shown in the figure, the average $^{13}$C', $^{15}$N, and $^{13}$C$^\alpha$ chemical shifts individually show dispersions of ~20.0 (=8.0*2.48), ~21.5 and ~53.7 (21.5 x 2.48) ppm, respectively, (where 1 ppm = 80 Hz, along $^{15}$N dimension at 800 MHz spectrometer) **(Fig. S1A, S1B and S1C)**. However, the linear combination of these chemical shifts provides dispersion relatively higher than the individual chemical shifts: (i) the subtraction and addition of $^{15}$N$_i$ and $^{13}$C'$_j$ chemical shifts (*i* and *j* represent the amino acid type) provide dispersion of 35.4 ppm and 35.5 ppm, respectively, **(Fig. S1D)** and (ii) the subtraction and addition of $^{15}$N$_i$ and $^{13}$C$^\alpha_j$ chemical shifts provide dispersion of 69.5 ppm and 68.0 ppm, respectively **(Fig. S1E)**. The fact has been exploited here in the form of reduced dimensionality experiments –(4,3)D-h<u>NCO</u>caNH and (4,3)D-h<u>N</u>co<u>CA</u>NH– to facilitate the assignment of backbone amide resonances of complex protein systems.

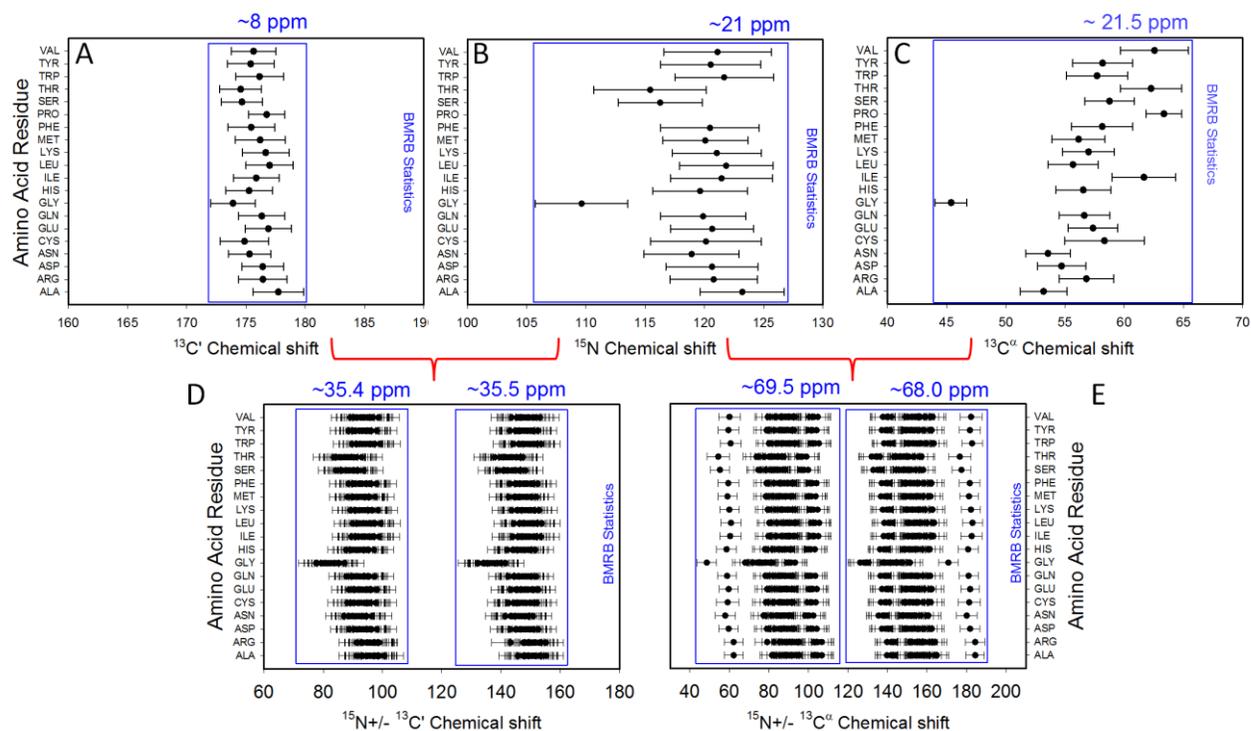

**Figure S1:** Comparison between dispersion of individual $^{15}N/^{13}C^{\alpha}/^{13}C'$ chemical shifts and that of linear combination of these chemical shifts for 20 common amino acids. In (A), (B) and (C) the average chemical shifts of $^{13}C'$, $^{15}N$, and $^{13}C^{\alpha}$ are plotted against the residue types. The average chemical shifts have been taken from the BMRB statistical table containing values calculated from the full BMRB database (http://www.bmrb.wisc.edu/ref_info/statful_.htm#1) [1]. This includes paramagnetic proteins, proteins with aromatic prosthetic groups, and entries where chemical shifts are reported relative to uncommon chemical shift references. Standard deviations in these values are plotted as error bars. In (D), the linear combinations of $^{15}N$ and $^{13}C'$ average chemical shifts (as evaluated according to **Eqns 1 and 2**) for all the amino acid types have been shown. In (E), the linear combinations of $^{15}N$ and $^{13}C^{\alpha}$ average chemical shifts (as evaluated according to **Eqns 1 and 2**) for all the amino acid types have been shown. The overall chemical shift dispersion achievable in each case has been shown at the top of each plot. Comparison clearly shows that the linear combination of chemical shifts lead to better dispersions compared to the individual chemical shifts.

## *Appendix-II*

***8M urea-denatured states of UNC60B from Caenorhabditis elegans*:**

The *Caenorhabditis elegans UNC60B* gene has been cloned into pETNH6 vector (Novagen) using restriction site *NcoI* and *BamHI* (in Dr. Ashish Arora's Lab, CDRI Lucknow). The cloning procedure added extra residues at N-terminal including a hexa-histidine tag and TEV-protease site. The clone was over-expressed in BL21 (λDE3) strain of *E. coli*. Conditions for optimal over-expression and purification were standardized. The yield of purified UNC-60B protein was 35 mg/L of culture medium. For isotopic labeling, over-expression of UNC-60B was standardized in minimal media containing $^{15}$N-ammonium sulfate and $^{13}$C-glucose (CIL, MA, USA) as the sole nitrogen and carbon sources, respectively. The purity of the sample was checked on SDS-PAGE. The protein sample was finally concentrated to ~800 µM using a centricon-filter (molecular weight cut off 3K, Amicon). For 8M urea-denatured state of the UNC60B, the concentrated protein (upto ~0.8 mM) was exchanged with 50 mM phosphate buffer of pH 6.0 (1 mM EDTA, 50 mM NaCl, 0.1% NaN$_3$, 1 mM DTT, and 8.0 M urea) containing 90% H$_2$O and 10% D$_2$O. The NMR experiments were started after keeping the solutions for about ~3 h so as to reach equilibrium.

**Table S1:** Acquisition parameters used to record the various spectra on the uniformly $^{15}N/^{13}C$ labeled Bovine apo-calbindin (75 aa), human ubiquitin (76 aa) protein and 8M urea denatured UNC60B.

| Parameters | Bovine Apo Calbindin-d9k/Human Ubiquitin | | 8M urea denatured UNC60-B | |
|---|---|---|---|---|
| | (4,3)D-h<u>NCO</u>caNH | (4,3)D-h<u>N</u>co<u>CA</u>NH | (4,3)D-h<u>NCO</u>caNH | (4,3)D-h<u>N</u>co<u>CA</u>NH |
| **Complex Data Points ($F_3$ X $F_2$ X $F_1$)** | 1024 X 36 X 64 | 1024 X 36 X 96 | 1024 X 40 X 72 | 1024 X 40 X 108 |
| **Spectral Width and (offset) in ppm** | $F_3(^1H)$ = 10.6 (4.7) $F_2(^{15}N)$ = 27 (118)/33(118) $F_1(NC')$ =110/100 (185) | $F_3(^1H)$ = 10.6 (4.7) $F_2(^{15}N)$ = 27 (118) $F_1(NC^\alpha)$ =160 (118/70) | $F_3(^1H)$ = 10.6 (4.7) $F_2(^{15}N)$ = 26 (118) $F_1(NC')$ =100 (118/185) | $F_3(^1H)$ = 10.6 (4.7) $F_2(^{15}N)$ = 26 (118) $F_1(NC^\alpha)$ =160 (118/70) |
| **No. of scans per FID (Recycle delay)** | 8 (0.8 sec) | 8 (0.8 sec) | 8 (0.8 sec) | 8 (0.8 sec) |
| **Zero filling** | 1024 X 128 X 256 | 1024 X 128 X 512 | 1024 X 128 X 256 | 1024 X 128 X 512 |
| **Total Expt. Time**≠ | ~23 hr, 11 min | ~1 day, 6 hr, 39 min | ~1 day, 1 hr, 33 min | ~1 day, 14 hr, 19 min |

*Note: A point to be mentioned here is that the proposed RD experiments require longer time compared to the normal 3D HN(C)N experiment in order to reach the same resolution. This is because of the fact that one has to acquire more data points along the co-evolved $F_1$ dimension where the linear combination of chemical shifts renders increased spectral width (generally two to three times of the primary $^{15}N$ spectral width).

## Appendix-III

### Amino Acid Sequence-Dependent Peak Patterns

Like the basic HN(C)N experiment, the self (notion used here for $H_iN_i$, $N_i^+$ and $N_i^-$) and sequential (notion used here for $H_{i+1}N_{i+1}$, $N_{i+1}^+$ and $N_{i+1}^-$) correlation peaks have opposite peak signs in different planes of the (4,3)D-hNCOcaNH and (4,3)D-hNcoCANH spectra except in special situations. Considering different triplets of residues, covering the general and all the special situations, the expected peak patterns in different planes of these spectra have been shown in **Figure S3** where self and sequential correlation peaks have been shown by squares and circles, respectively. **Figure S3A** shows the expected peak patterns (for the example stretch -PZXGYP- covering all the various possibilities) in the $F_2(^{15}N)$-$F_3(^1H)$ planes at $F_1 = N_{i+1}^+ / N_{i+1}^-$ chemical shift identified for residue, $i$. **Figure S3B** shows the expected peak patterns (for the triplet stretches shown above each panel) in the $F_1(NC)$-$F_3(^1H)$ planes of these spectra at the $F_2(^{15}N)$ chemical shift of the central residue, $i$. The filled and empty circles/squares represent positive and negative peaks, respectively. The actual signs in the spectrum are dictated by whether the $i-1$ residue is a glycine or otherwise, and of course by the phasing of the spectrum. Here, the self-correlation peaks (i.e. $H_iN_i$, $N_i^+$ and $N_i^-$) have been made positive (like the basic HN(C)N spectrum) for a triplet sequence –XYZ-, where X, Y, and Z are any residues other than glycines and prolines which have been represented here by letters "G" and "P", respectively. Accordingly the peak patterns have been shown in **Figure S3**. From the figure, it is clear that glycines make an important difference in the expected patterns of peaks. For glycines, the evolutions of the magnetization components are slightly different because of the absence of the $\beta$-carbon. This in turn generates some special patterns depending on whether the $(i-1)^{th}$ residue is a glycine or otherwise (see **Fig. S3**). These special peak patterns –which help in the identification of residues following glycines in the sequence– provide important start or check points during the course of sequential assignment process. Prolines which do not have amide proton further give rise to special patterns in the spectrum. Prolines, at position $i$-1 in the triplet stretches, results in the absence of sequential amide correlation peaks ($H_{i-1}$, $N_{i-1}$) in $F_2(^{15}N)$-$F_3(^1H)$ plane of the spectrum at $F_1 = N_i^+ / N_i^-$ (see **Fig. S3A**), whereas a proline, at position $i$+1 in the triplet stretches results in the absence of correlation peaks ($H_i, N_{i+1}^+$ and $H_i, N_{i+1}^-$) in $F_1(NC)$-$F_3(^1H)$ plane of the spectrum at $F_2 = ^{15}N_i$ (see **Fig. S3B**). These special peak patterns –which help in the identification of residues following prolines in the sequence–, provide important stop/break points during the course of the sequential assignment walk and thus facilitates the assignment process further.

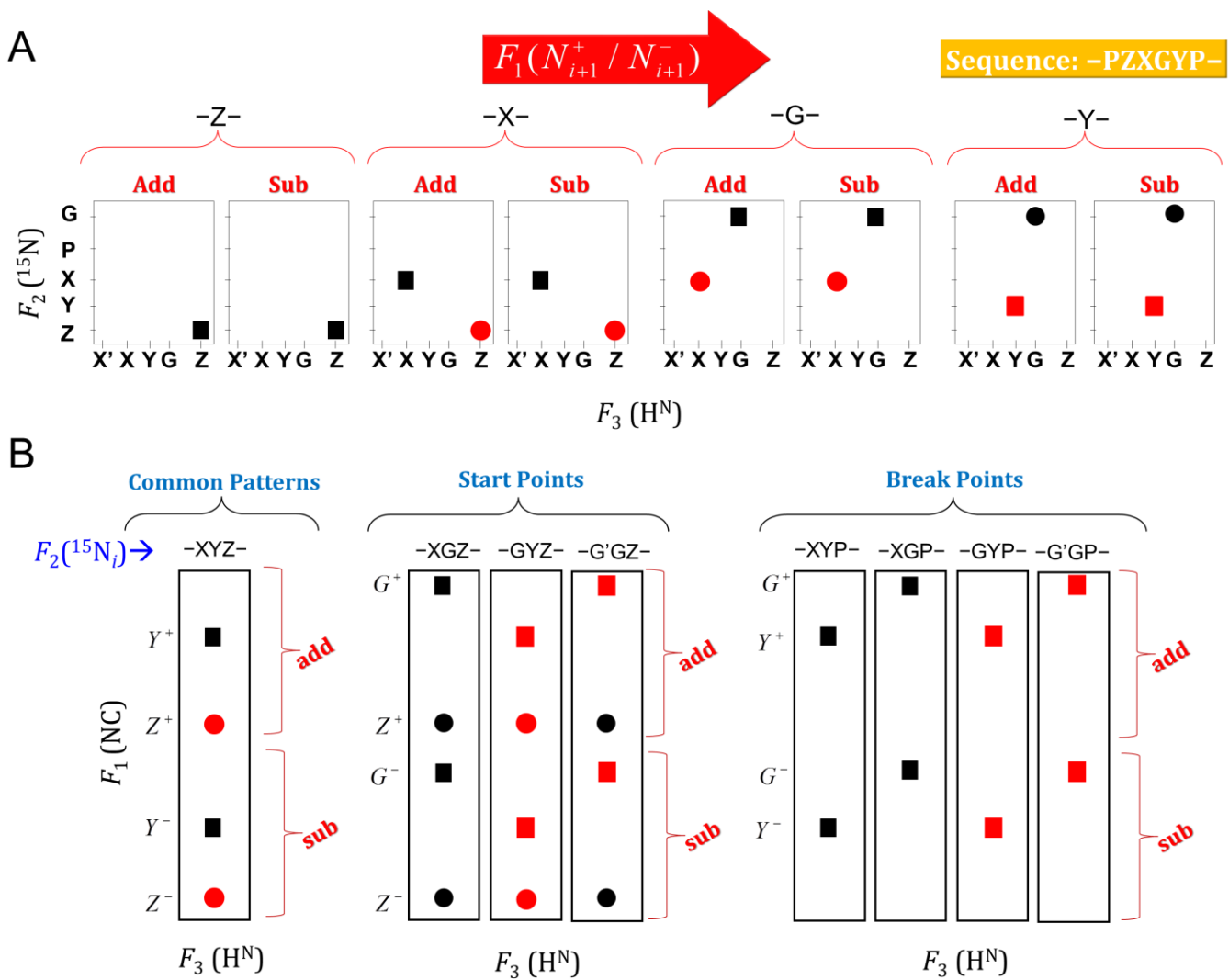

**Figure S2: (A)** Schematic peak patterns in the $F_2(^{15}N)$-$F_3(^1H)$ planes of the (4,3)D-hNCOcaNH and (4,3)D-hNcoCANH spectra at $F_2 = N^+_{i+1} / N^-_{i+1}$. An arbitrary amino acid sequence –PZXGYP– (covering all the common and special peak patterns) is chosen to illustrate the triplet specific peak patterns in these planes of the spectra. In the sequence, X, Y and Z represent any residue other than glycine and proline. G and P, respectively, represent the glycine and proline residues. Black and red colors represent positive and negative signs, respectively. Squares and circles represent the self and sequential peaks, respectively. **(B)** Schematic peak patterns in the $F_1$(NC)-$F_3(^1H)$ planes and the corresponding triplet of residues. The patterns involving G serve as start and/or check points during the sequential assignment walk through the spectrum. GGP, which is likely to be less frequent than the others, may serve as a unique start point. The patterns involving P identify break points during a sequential walk and thus they also serve as check points. The pattern, which does not involve a G or a P, is the most common one occurring through the sequential walk.

**Figure S3: (A)** The illustrative stretch of sequential walk through the strips along $F_1$(NC) dimension and centered about amide $^1$H chemical shift of residue $i$ in $F_1$(NC)-$F_3$($^1$H) planes of the (4,3)D-h<u>NCO</u>caNH spectrum of human ubiquitin (for residues Lys6-Ile13) at $^{15}$N chemical shift of residue shown at the top of a particular strip. **(B)** The illustrative stretch of sequential walk through the strips along $F_1$(NC) dimension and centered about amide $^1$H chemical shift of residue $i$ in $F_1$(NC)-$F_3$($^1$H) planes of the (4,3)D-h<u>N</u>co<u>CA</u>NH spectrum of bovine apo calbindin-d9k (for residues Leu6-Tyr13) at $^{15}$N chemical shift of residue shown at the top of each individual strip. Each strip contains information about self ($N_i^+$ and $N_i^-$) and sequential ($N_{i+1}^+$ and $N_{i+1}^-$) correlation peaks. The red and black contours indicate positive and negative peaks, respectively. The labels and numbers at the top in each panel identify the residue and the respective $F_2$($^{15}$N) chemical shifts. A horizontal line connects a sequential peak (red here) in one plane to the self peak (black here) in adjacent plane on the right. However, the special patterns appear for strips of glycines where both self and sequential peaks appear positive/black in sign e.g. Gly10 and Gly8 strips in **(A)** and **(B)**, respectively. On the other hand, self and sequential peaks appear negative/red in sign on the strips corresponding to residues following glycines in the sequence e.g. Lys11 and Ile9 strips in **(A)** and **(B)**, respectively.

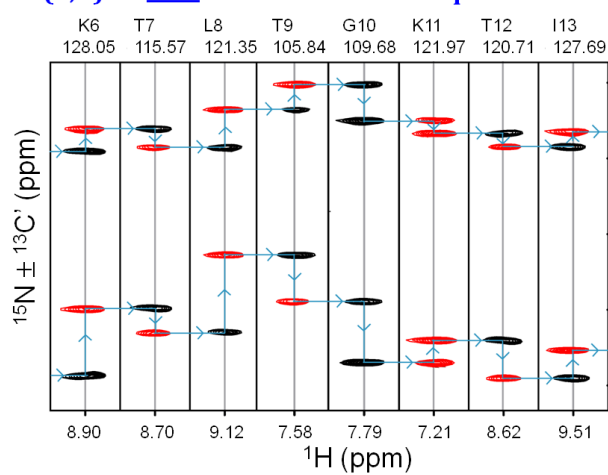

**A: (4,3)D h<u>NCO</u>caNH of human Ubiquitin**

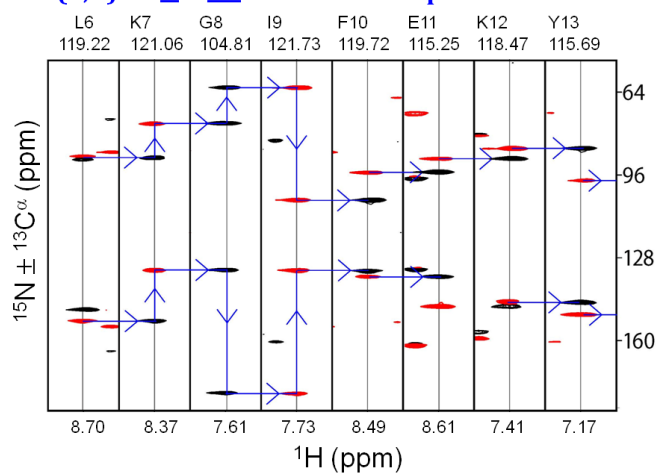

**B: (4,3)D h<u>N</u>co<u>CA</u>NH of bovine apo calbindin**

**Figure S4:** An illustrative example showing use of correlations observed in the $F_2$($^{15}$N)–$F_3$($^1$H) planes at the $^{15}N^+_{i+1}$ / $^{15}N^-_{i+1}$ chemical shifts identified for residue *i* for establishing a sequential (*i* → *i*+1) connectivity between the backbone amide correlation peaks of $^1$H-$^{15}$N HSQC spectrum (like in case of HN(C)N based assignment protocol [2]). Red and black contours represent positive and negative phase of the peaks, respectively. A sequential amide cross peak (H$_{i+1}$, N$_{i+1}$) exists in both the $F_2$($^{15}$N)–$F_3$($^1$H) planes of the spectrum at the $^{15}N^+_{i+1}$ and $^{15}N^-_{i+1}$ chemical shifts identified for the residue *i*. The other advantage is the opposite peak signs for the self (H$_i$, N$_i$) and sequential (H$_{i+1}$, N$_{i+1}$) amide correlation peaks. These spectral features help to reduce the search for the sequential amide correlations rather than making the search in various planes of the 3D spectrum (as is the case with other presently used assignment strategies).

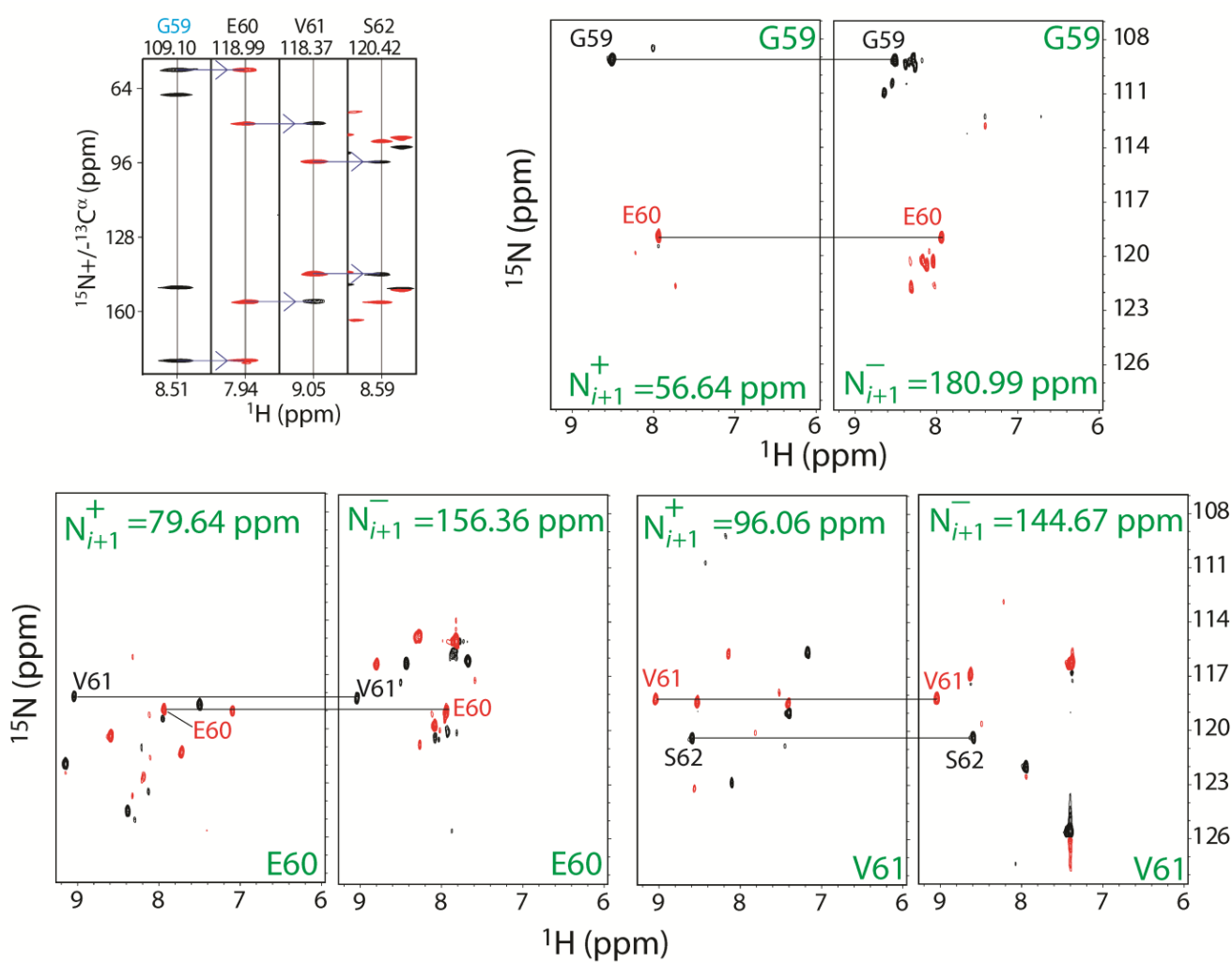

**Figure S5:** *F*$_1$-*F*$_3$ projection planes, respectively, of (4,3)D-H<u>NCO</u>caNH **(A)** and (4,3)D-h<u>N</u>co<u>CA</u>NH **(B)** spectra of chicken SH3 domain projected along the *F*$_2$($^{15}$N) axis. As described earlier in [3], the sequential assignment walk can be performed directly on these planes for small to medium sized well folded proteins. The details will be presented elsewhere.

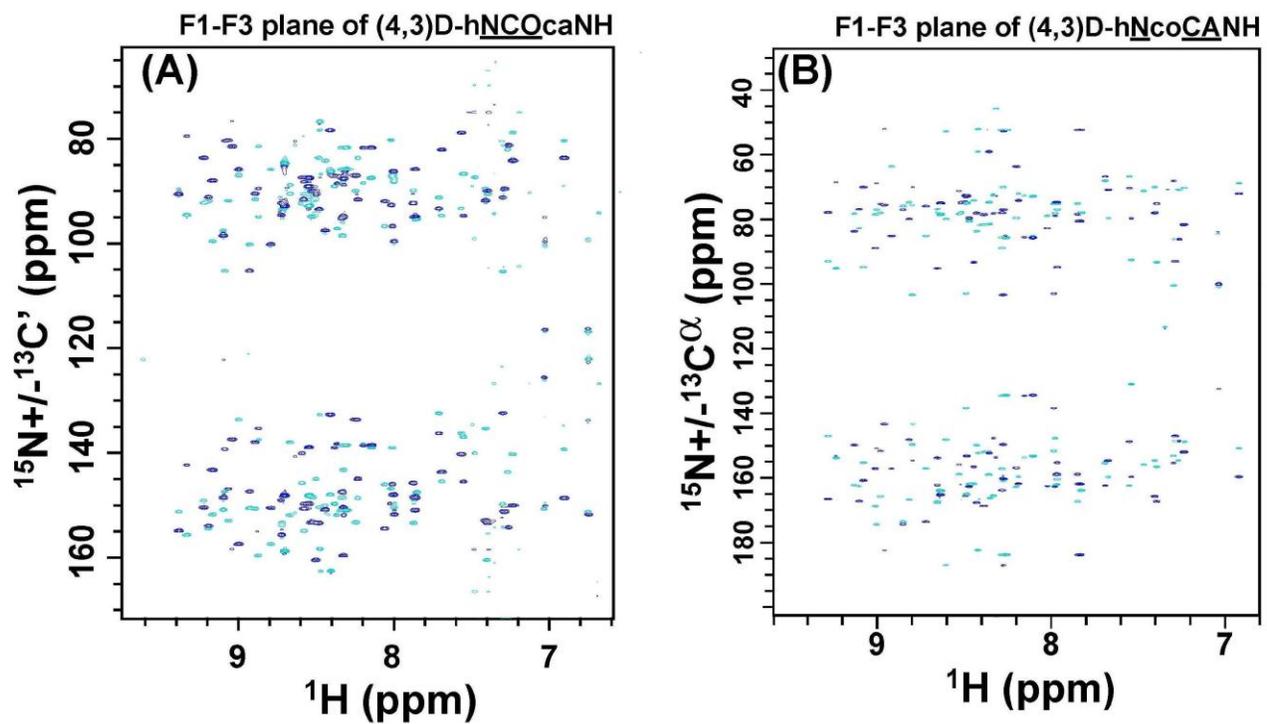

**Figure S6: (A)** $^1$H-$^{15}$N HSQC spectrum of UNC60B in 8 M Urea at pH 6.0 and 298 K. **(B)** Sequential walk through the $F_1$-$F_3$ planes of the (4,3)D-h<u>NCO</u>caNH spectrum. Sequential connectivities are shown for Thr36 to Val39 stretch. **(C)** Sequential walk through the $F_1$-$F_3$ planes of the (4,3)D-h<u>Nco</u><u>CA</u>NH spectrum of UNC60B. Sequential connectivities are shown for Val5 to Asp8 stretch. The $F_2$ ($^{15}$N) values are shown at the top for each strip. The black and red contours indicate positive and negative peaks, respectively. **(D)** The summary of sequential assignment obtained along the amino acid sequence of UNC60B. The residues underlined have been assigned here using the RD (4,3)D-h<u>Nco</u><u>CA</u>NH experiment (for establishing the sequential connectivities) in combination with routine 3D-CBCAcoNH experiment (for amino acid type identification). Gly shown in red served as the start/check point while the Pro shown in blue served as the break point.

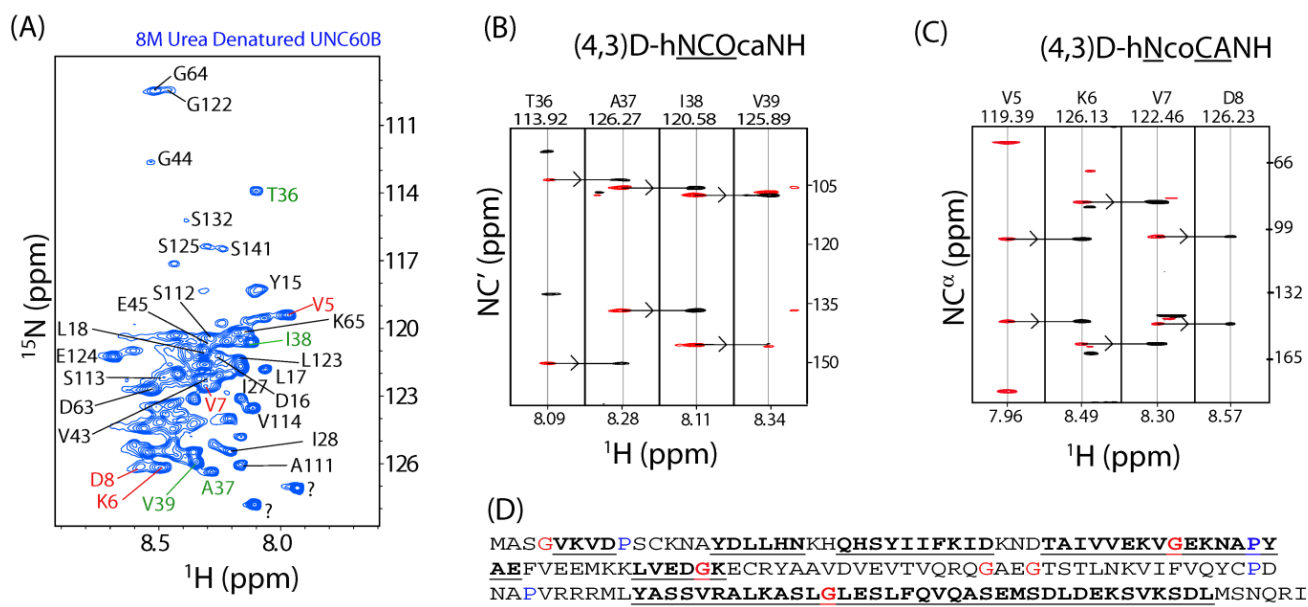

*Kumar D*

## Reference:

[1] BMRB, Statistics Calculated for Selected Chemical Shifts from Atoms in the 20 Common Amino Acids (http://www.bmrb.wisc.edu/ref_info/), (2011)

[2] A. Chatterjee, N.S. Bhavesh, S.C. Panchal, R.V. Hosur, A novel protocol based on HN(C)N for rapid resonance assignment in ((15)N, (13)C) labeled proteins: implications to structural genomics, Biochem. Biophys. Res. Commun. 293 (2002) 427-432.

[3] D. Kumar, A. Borkar, R.V. Hosur, Facile Backbone (1H, 15N, 13Ca and 13C') Assignment of 13C/15N labeled proteins using orthogonal projection planes of HNN and HN(C)N experiments and its Automation, Magn Reson. Chem. 50 (2012) 357-363.

*13*